\documentclass[aps,prl,superscriptaddress,twocolumn]{revtex4-1}
\usepackage[T1]{fontenc}
\usepackage[latin1]{inputenc}
\usepackage{amsmath,amsfonts,amssymb}
\usepackage{wrapfig}
\usepackage{graphicx}
\usepackage{bbm}
\usepackage{rotating}
\usepackage[tight]{units} 
\usepackage{nicefrac}
 	
\newenvironment{methods}{
    \section*{Methods}
    \setlength{\parskip}{12pt}
    }{}
\newenvironment{addendum}{
        \setlength{\parskip}{12pt}
    }{}

\begin{document}



\title{Trapped-ion antennae for the transmission of quantum information}
\author{M. Harlander}
\affiliation{Institut für Experimentalphysik, Universität Innsbruck, Technikerstrasse 25, A-6020 Innsbruck, Austria}

\author{R. Lechner}
\affiliation{Institut für Experimentalphysik, Universität Innsbruck, Technikerstrasse 25, A-6020 Innsbruck, Austria}

\author{M. Brownnutt}
\affiliation{Institut für Experimentalphysik, Universität Innsbruck, Technikerstrasse 25, A-6020 Innsbruck, Austria}

\author{R. Blatt}
\affiliation{Institut für Experimentalphysik, Universität Innsbruck, Technikerstrasse 25, A-6020 Innsbruck, Austria}
\affiliation{
Institut für Quantenoptik und Quanteninformation, Österreichische Akademie der Wissenschaften,
Otto-Hittmair-Platz 1/Technikerstraße 21a, A-6020 Innsbruck, Austria}

\author{W. Hänsel}
\altaffiliation{Present address: Menlo Systems GmbH, Am Klopferspitz
19a, D--82152, Martinsried, Germany}
\affiliation{Institut für Experimentalphysik, Universität Innsbruck, Technikerstrasse 25, A-6020 Innsbruck, Austria}
\affiliation{
Institut für Quantenoptik und Quanteninformation, Österreichische Akademie der Wissenschaften,
Otto-Hittmair-Platz 1/Technikerstraße 21a, A-6020 Innsbruck, Austria}
\date{\today}

\begin{abstract}
\end{abstract}

\maketitle
\noindent \textbf{More than one hundred years ago Heinrich Hertz succeeded in transmitting signals over a few
meters to a receiving antenna using an electromagnetic oscillator and thus proving the electromagnetic theory developed by James C. Maxwell\cite{Hertz1890}.
Since then, technology has developed, and today a variety of oscillators is
available at the quantum mechanical level. 
For quantized electromagnetic oscillations atoms in cavities can be used to couple electric fields\cite{Kuhr2007,Gleyzes2007}.
For mechanical oscillators realized, for example, with
cantilevers\cite{Anetsberger2009,Schliesser2006} or vibrational modes of
trapped atoms\cite{Kinoshita2006} or ions\cite{Leibfried2003,Monroe1996}, a quantum
mechanical link between two such oscillators has, to date,
been demonstrated in very few cases and has only been achieved in
indirect ways. Examples of this include the mechanical transport
of atoms carrying the quantum information\cite{Jost2009} or the
use of spontaneously emitted photons\cite{Blinov2004}. In this
work, direct coupling between the motional dipoles of separately
trapped ions is achieved over a distance of 54\,$\boldsymbol\mu$m,
using the dipole-dipole interaction as a quantum-mechanical
transmission line\cite{Cirac2000}. This interaction is small
between single trapped ions, but the coupling is amplified by
using additional trapped ions as antennae. With three ions in each
well the interaction is increased by a factor of seven as compared
to the single-ion case. This enhancement facilitates bridging of
larger distances and relaxes the constraints on the
miniaturization of trap electrodes. This represents a new building
block for quantum computation and also offers new opportunities to
couple quantum systems of different natures.}

\noindent The exchange of quantum information between qubits at remote sites
is a key feature required to render quantum computation truly
scalable\cite{DiVincenzo2000}. The dipole-dipole interaction offers a link between
separate quantum systems without the need to shuttle particles
between sites. The interaction strength depends on the orientation
and distance of the dipoles and is in general given by
\begin{equation}
U_{\mathrm{dd}} = \frac{1}{4 \pi \epsilon_{0}}\frac{
\vec{d_{1}}\vec{d_{2}}-3(\vec{d_{1}}\vec{e_{\mathrm{r}}})(\vec{d_{2}}\vec{e_{\mathrm{r}}})}{r^{3}}
\,, \label{eq:dipole-dipole}
\end{equation}
where $d_{i}$ are the interacting dipoles, $r$ and $\vec{e}_r$ denote the magnitude and direction of their separation.

\begin{figure}[!ht]
\begin{center}
\includegraphics[width=0.4\textwidth]{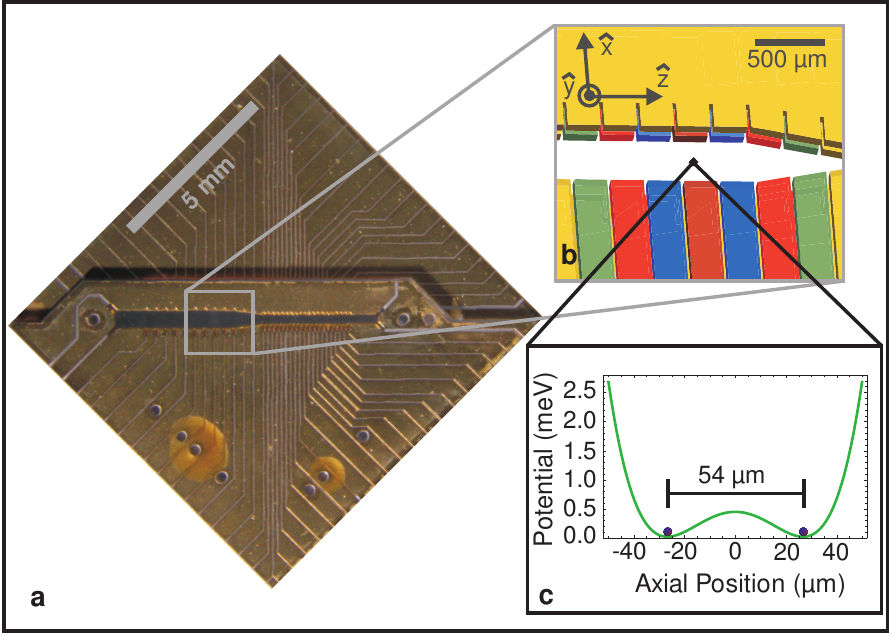}
\caption{\textbf{(a)} Image of the two-layer segmented trap with gold-on-alumina electrodes. 
The electrodes are 250\,$\mathrm{\mu m}$ wide with a pitch of 280\,$\mathrm{\mu m}$ and the ion-electrode spacing is 258\,$\mathrm{\mu m}$. \textbf{(b)} Schematic close-up of the zone of
interest. The electrode pairs used to generate the double-well potential are colour-coded: red for positive voltages, blue for negative voltages and green for compensation voltages. \textbf{(c)} The axial potential calculated with the parameters determined from the motional spectra in Fig.\,\ref{fig:CoupledSpectra} is illustrated.\label{fig:TrapAndPotential}}
\end{center}
\end{figure}
\noindent Here this interaction is explored using ions or ion strings held
in two separate potential wells of a linear segmented ion trap
(see Fig.\,\ref{fig:TrapAndPotential}), where the interacting dipoles are produced
by the oscillating charges.
 As the dipole-dipole interaction decreases rapidly
with trap separation, it is advantageous to bring the trapping
wells as close together as possible. However, the generation of
small inter-well distances requires similarly small distances, $d$,
between the ions and the trap electrodes. This requirement runs
counter to the effort to keep a larger ion-electrode separation in
order to reduce ``anomalous
heating''\cite{Deslauriers2006,Seidelin2006} with its $d^{-4}$
scaling, and also the effects of technical noise of the applied
voltages\cite{Leibrandt2009}. Various routes may be taken to
balance these competing requirements.
In one approach, the heating rate may be reduced through the use
of cryogenic temperatures\cite{Labaziewicz2008}.
Here, another approach is taken that
uses more ions in the individual traps and enables interaction
over larger distances.
The additional ions work as ``antennae'' that increase the
motional dipole moment at the respective trapping site.

\noindent Given longitudinal alignment of the traps with one particle in
each well, the dipole-dipole interaction
(Eq.\,\ref{eq:dipole-dipole}) is:
\begin{eqnarray}
 U_{\mathrm{dd}}&=&-\frac{q_{1} q_{2}}{2 \pi \epsilon_{0}}
 \frac{\Delta z_{1}\Delta z_{2}}{r^3} \\
 \nonumber
 &=& - \hbar \frac{\Omega_{\mathrm{c}}}{2}
 (a_{1}+a_{1}^\dagger)(a_{2}+a_{2}^\dagger) \\
 &\approx& - \hbar \frac{\Omega_{\mathrm{c}}}{2} \left(a_{1} a_{2}^{\dagger} +
 a_{1}^{\dagger}a_{2}\right) \label{eq:ExchangeInteraction}
\end{eqnarray}
with
\begin{equation}
 \Omega_{\mathrm{c}}=\frac{q_{1}q_{2}}{2\pi\epsilon_0 \sqrt{m_1 m_2\, \omega_1 \omega_2}}\frac{1}{r^3}\,. \label{eq:CouplingStrength}
\end{equation}
Here $q_i$ and $m_i$ refer to the charge and mass of the
particles, $\Delta
z_i=\sqrt{\frac{\hbar}{(2 m_i\omega_i)}}(a_i+a_i^\dagger)$ denotes the
vibrational amplitude of the motion of ion $i$. The quantum-mechanical creation and annihilation
operators, $a_i$ and $a_i^\dagger$ act on the
individual harmonic oscillators with frequencies $\omega_i$. Rapidly 
oscillating terms have been neglected in Eq.\,3.

At resonance, i.\,e. for $\omega_1=\omega_2$, the coupling
described by Eq.\,(\ref{eq:ExchangeInteraction}) leads to a
complete exchange of motional states between the two ions after
time $T_{\mathrm{swap}}=\pi/\Omega_{\mathrm{c}}$. This
is analogous to two coupled pendula connected by a (massless)
spring. If one pendulum initially oscillates while the other is at
rest, the motion is periodically exchanged between them. The first
pendulum comes to a complete stop after a characteristic time,
$T_\mathrm{swap}$, proportional to the associated spring constant.
A quantum-mechanical description may be given using the
vibrational quanta, $n_i$, often labelled ``phonons'', in the individual wells. Under
the dipole-dipole interaction an initial motional state
$\left|n_1,n_2\right\rangle$ becomes entangled with all other motional states
$\left|n'_1,n'_2\right\rangle$ respecting $n'_1+n'_2=n_1+n_2$.
Only at odd (even) multiples of $T_{\mathrm{swap}}$ is the swapped
(original) basis state recovered.
Notably, the initial state $\left|0,1\right\rangle$ evolves into
the Bell-state $(\left|0,1\right\rangle +
\left|1,0\right\rangle)/\sqrt 2$ after time $T_{\mathrm{swap}}/2$,
yielding a maximally entangled state of motion. This motional
 entanglement can be mapped on to the internal electronic state of the ions\cite{Cirac1995}.

\begin{figure}
\begin{center}
\includegraphics[width=0.45\textwidth]{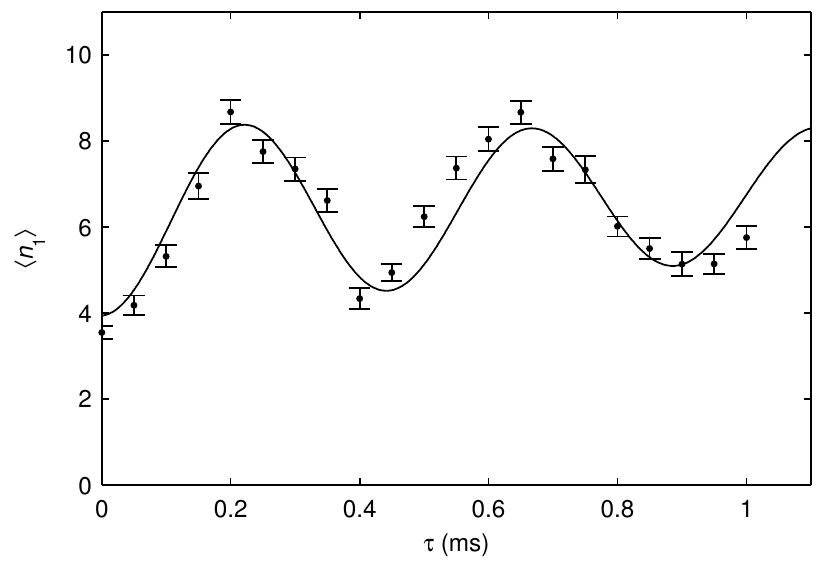}
\caption{Energy exchange between two trapped ions over a
distance of 54\,$\mu$m. The data show the average number of
phonons in the first trapping well, $\langle n_{\mathrm{1}}\rangle$, after sideband cooling and a variable waiting time, $\tau$. The sequence for sideband cooling has been
arranged to yield different phonon numbers in the two wells,
revealing the state exchange as an oscillation of phonon numbers
at the level of a few quanta.
The observed time for a complete exchange is $T_{\mathrm{swap}}= 222(10)\,\mu$s, indicating a mode splitting of
$\Omega_{\mathrm{c}}\approx 2\pi\times 2.25(4)$\,kHz. A damping constant
$\tau_{\mathrm{damp}}$=3(2)\,ms and a constant background heating of 1.3(7) quanta per millisecond are
inferred from the fit to the data. The error bars indicate one standard deviation
as inferred from Monte Carlo simulations. Lateral deviations of the data from the fit are attributed to small drifts in the resonance condition which modify the exchange rate.} \label{fig:ExchangeWithFit}
\end{center}
\end{figure}

\noindent In the experiment presented the coherent energy exchange is
demonstrated between singly charged $\mathrm{^{40}Ca^{+}}$ ions.
They are held in an ion trap with gold-on-alumina electrodes
arranged in a two-layer geometry (cf.
Fig.\,\ref{fig:TrapAndPotential}) similar to the one described in Ref.\,18. Applying DC
voltages of up to \unit[110]{V} to seven adjacent electrode pairs, a double-well potential with a trap separation of
$\unit[54]{\mathrm{\mu m}}$ and trap frequencies of
\unit[537]{kHz} is created (see Methods).
The ions are Doppler-cooled on the S$_{1/2}$-P$_{1/2}$ transition
at 397\,nm using a single, elliptically shaped laser beam, and
detected by collecting the fluorescence light on an EM-CCD camera
and on a photo-multiplier tube. Two \unit[729]{nm} laser beams,
individually focused on the two trapping sites, are used to
perform sideband cooling on the S$_{1/2}$-D$_{5/2}$ transition and
to map out the sideband spectrum\cite{Roos1999}.
During the \unit[4]{ms} period of sideband cooling the
\unit[729]{nm} beam is alternated between the two trapping sites
at the approximate rate of the energy exchange, leading to an
imbalance of phonon population between the traps. Immediately
following the cooling cycle a mean phonon number
$\left<n_1\right>=3.9(4)$ is observed in the first well, by
comparing Rabi-flops on the red- and blue-sideband\cite{Roos1999}.
Fig.\,\ref{fig:ExchangeWithFit} shows
the oscillatory behaviour of the phonon number in this well as a
function of waiting time. The theoretical fit to the data
indicates an exchange within $T_{\mathrm{swap}}=222(10)\,\mu$s and an initial
phonon population of $\left<n_2\right>=9(1)$ in the second
well.
Currently, the ions experience a heating rate of $1.3(7)$
quanta per millisecond, which is comparable to other
room-temperature traps\cite{Turchette2000}.

\noindent Ideally, the exchange rate would be significantly larger than the
average heating rate. To enhance the exchange rate
strings of several ions can be used in the trapping wells.
Approximating the individual ion strings as point-like objects
with appropriately increased charge and mass, the resulting
dipoles scale with the square root of the number of ions (see
Eq.\,\ref{eq:CouplingStrength}), leading to an overall increase of
the interaction rate proportional to the total number of ions.
Using lighter ion species for antennae would enhance the coupling even more.

\begin{figure}
\begin{center}
\includegraphics[width=0.45\textwidth]{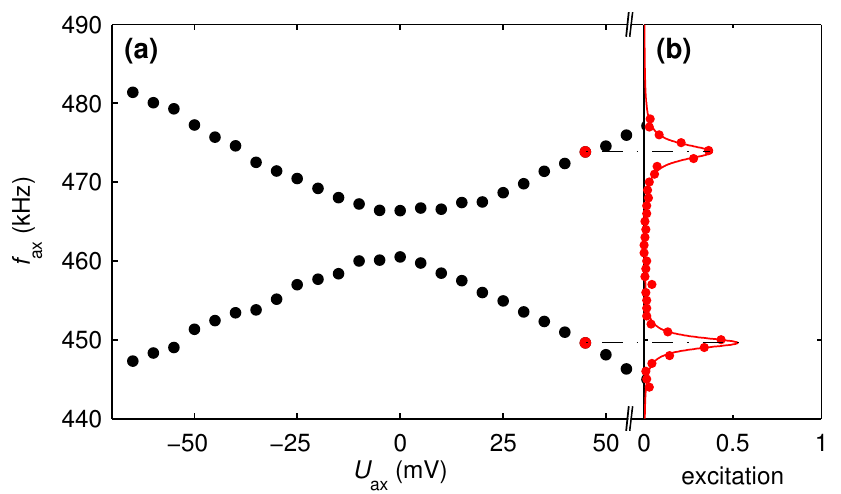}
\caption{\textbf{(a)} Oscillation frequencies of
four trapped ions (two in each well) as a function of the axial
control voltage $U_{\mathrm{ax}}$, yielding a mode splitting of
$5.5(3)$\,kHz. The data points correspond to peaks in individual
sideband spectra taken on the S$_{1/2}$-D$_{5/2}$ transition. The error bars are smaller than the dot size. \textbf{(b)} Example of an individual sideband spectrum.}
\label{fig:Coupling2x2}
\end{center}
\end{figure}

\begin{figure}
\begin{center}
\includegraphics[width=0.45\textwidth]{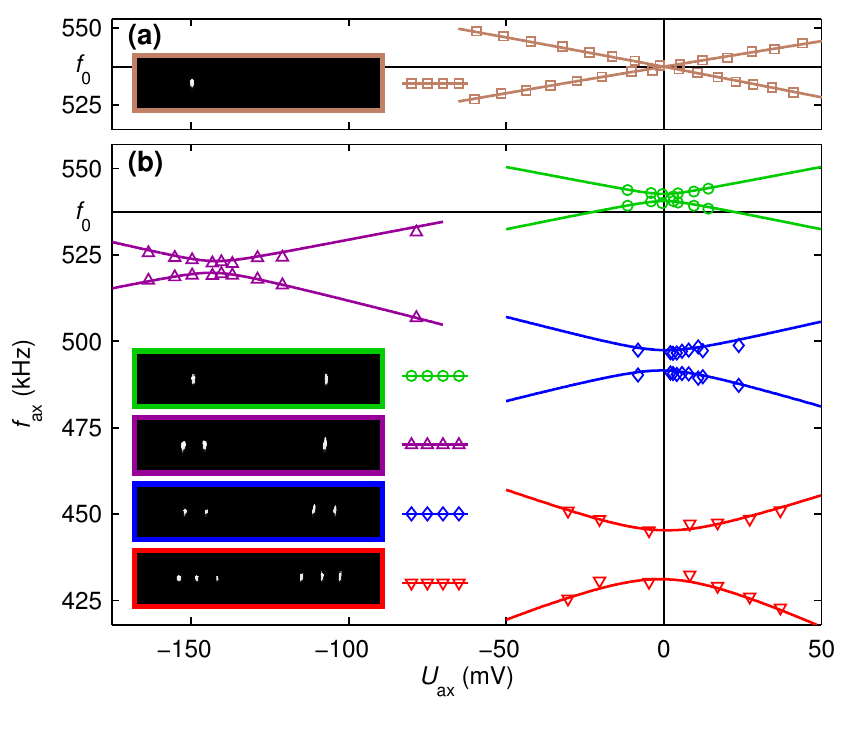}
\caption{Experimentally observed dipole-dipole coupling for
various ion configurations in a double-well potential. The graphs
display the oscillation frequencies of the two lowest vibrational
modes as a function of control voltage $U_\mathrm{ax}$ and reveal
the mode splitting at resonance. \textbf{(a)} Unmodified
trap frequencies, as measured with a single ion, in the first or
second well, respectively.
\textbf{(b)} Spectra with up
to three ions in each well as depicted in the insets. The lines
represent predictions from numerical calculations and fit all data
simultaneously. A small drift in the control voltage of
$\unit[7]{\mathrm{mV}}$ over one hour has been taken into
account, aligning the spectra on top of one another. By the
use of three ions in each potential well the coupling is increased
from 1.9(3)\,kHz to 14(1)\,kHz (see text).
The vertical elongation of the ion pictures is due to abberations, 
arising from an off-axis position of the ions relative to the imaging axis.
} \label{fig:CoupledSpectra}
\end{center}
\end{figure}

\noindent This increase is demonstrated by mapping out the ions' excitation
spectrum as the trapping frequencies of the individual sites are
scanned through the resonance condition. The frequency scan is
achieved by applying a control voltage $U_\mathrm{ax}$ to an outer
trap-electrode pair (see Methods). The dipole-dipole coupling
manifests as an avoided crossing, separating the mode frequencies
by $\Omega_\mathrm{c}$. Close to resonance, the motion of the ion
strings is strongly coupled, and the oscillation can be excited
with \unit[729]{nm} light on either of the two trapping sites\cite{Roos1999}.
Fig.\,\ref{fig:Coupling2x2} \textbf{(a)} shows an example of this 
avoided crossing measured with two ions in each well 
while Fig.\,\ref{fig:Coupling2x2} \textbf{(b)} represents an individual 
sideband spectrum.
\noindent The motional spectra of five ion configurations have been
analyzed, using up to three ions in each well. The configurations
and corresponding mode spectra are displayed in
Fig.\,\ref{fig:CoupledSpectra}. The curves present calculations
using a common set of fit parameters for the potential and for the
action of the control voltage $U_\mathrm{ax}$ (see Methods). 
\noindent For the configuration with one ion per well, the observed splitting of
\unit[1.9]kHz agrees within one standard deviation with the energy-exchange
rate from Fig.\,\ref{fig:ExchangeWithFit}, which was measured with
the same trap parameters. The data further show that the splitting
is increased from $\unit[1.9(3)]{\mathrm{kHz}}$ to
$\unit[14(1)]{\mathrm{kHz}}$ by using up to three ions in each
well, without any modification of the external potential. The
seven-fold increase of the coupling is beyond the factor three that is
expected from a simple point-charge model and is due to the
anharmonicity of the individual potential wells: as the outer
potential walls are steeper than the inner ones, the ion-strings'
centers of mass get closer as more ions are added. At the same
time, the average oscillation frequency is reduced. Both of these
effects lead to an increase in the coupling strength (see
Eq.\,\ref{eq:CouplingStrength}). The extent of the ion strings
provides an additional increase when it becomes non-negligible
with respect to the inter-well distance.

\noindent The demonstrated coupling\cite{Brown2010} can be used in diverse schemes to create
entanglement or to perform gates. The creation of Bell-states is
discussed above and requires ground state cooling of the ions. One
route to perform a thermally robust quantum gate has been
described by Cirac et al.\cite{Cirac2000}, however it requires a
state-dependent pushing force by an appropriate laser. It should
be noted that a fast variant\cite{Benhelm2008} of the thermally
robust M{\o}lmer-S{\o}rensen gate\cite{Sorensen1999} which uses bichromatic
 laser beams can also be implemented in a dipole-dipole coupled system 
 between different trapping sites. If the laser frequencies are
tuned to the mid-points of the red and blue sideband pairs, a gate
can be performed within $T_\mathrm{gate}=4\pi/\Omega_\mathrm{c}$.
The new coupling scheme also has a great impact on possible
architectures for quantum computation. It is straight forward to
imagine a linear array of ion traps, each holding a string of
ions. Moreover, the coupling of parallel dipoles differs from longitudinal coupling by only a
factor of $-1/2$, while under the quadrupole angle of
$\theta_q=\arccos(\sqrt{1/3})$ the coupling vanishes. Considering oscillations along the
trap axis, a two-dimensional array of traps 
may thus be realized, where unwanted coupling to nearest
diagonal neighbours is suppressed. Bringing rows of traps pairwise into resonance 
allows the creation of linear cluster states within two steps. A
two-dimensional cluster state would only require two further steps by pairwise coupling of columns\cite{Raussendorf2003}.

\noindent The amplification of the dipole-dipole interaction is not limited
to ions.
The presented technique may be directly transferred to the
coupling of trapped Rydberg atoms\cite{Urban2009}.
Furthermore, atomic and ionic systems may be
combined. Neutral atoms have already been brought close to ions\cite{Zipkes2010,Schmid2010}. When the ions' oscillation frequency
is tuned into resonance with adjacent levels of a Rydberg atom,
coupling between ions and Rydberg atoms may become achievable.

\begin{methods}
\subsection{Calculation of eigenfrequencies in the double-well potential.}
\noindent A symmetric double-well potential is created along the trap axis
by applying \unit[2.8]{V}, \unit[110.4]{V}, \unit[-16.8]{V},
\unit[13.7]{V}, \unit[-16.8]{V}, \unit[110.4]{V}, and
\unit[-33.0]{V} to the seven adjacent electrodes pairs marked in Fig.\,\ref{fig:TrapAndPotential}\textbf{(b)}. The
outermost electrode pairs are used to compensate for the
asymmetry induced by the taper zone. 
The control voltage, $U_\mathrm{ax}$, is additionally applied to the leftmost electrode pair, 
creating a nearly-homogeneous electric field in the center region. 
This shifts the trap frequencies of the two wells in opposite directions and 
is used to tune the wells into and out of resonance. 
In the region of interest the double-well potential
is well described by a fourth-order polynomial, where the 0th
order term does not influence the dynamics and the third-order
coefficient has been eliminated by appropriate choice of origin:
\begin{equation}
 U_{\mathrm{pot}}(z)= \alpha_1\cdot z +\alpha_2\cdot z^2+\alpha_4\cdot z^4 \,.
\end{equation}
The polynomial coefficients $\alpha_2$ and $\alpha_4$ determine
both the uncoupled trap frequency, $\omega_0$, and the inter-well
separation, $r$, at the symmetry point\cite{Home2006} ($\alpha_1=0$).
For the theoretical calculations, the action of the tuning voltage is modelled 
as a linear contribution to the coefficients $\alpha_1=\mathrm{\unit[4.61\times10^{-24}]{m^{-1}}}$ 
and $\alpha_2=\mathrm{\unit[0.9\times10^{-20}]{m^{-2}}}$.
The equilibrium positions
for $N$ ions, $\vec{z}_0^N=(z_{0,1},...,z_{0,N})$, are those
minimizing the total energy, $U_{tot}^N$, including the Coulomb
interaction:
\begin{equation}
U_{\mathrm{tot}}^N (\vec{z}^N)=\sum_{i=1}^{N}U_{\mathrm{pot}}(z_i)
+ \sum_{i=1}^{N-1}\sum_{j=i+1}^{N} \frac{q^2}{4 \pi
\epsilon_0}\cdot\frac{1}{\left| z_i - z_j \right|}\,.
\end{equation}
The theoretical data for the vibrational mode frequencies,
$\omega_i$, in Fig.\,\ref{fig:CoupledSpectra} are computed from
the two lowest eigenvalues of the Jacobian matrix of the total
energy $U_{\mathrm{tot}}^N(\vec{z}^N)$ at positions $\vec{z}_0^N$.
These correspond to the modes where the individual ion strings
move as a whole, approximating the motion of single particles.
\end{methods}

\begin{addendum}
\section{Acknowledgements}
\noindent We thank Hartmut H\"{a}ffner for fruitful discussions at an early state of the project. 
We gratefully acknowledge the support of the EU STREP project MICROTRAP, the Austrian Science Fund (FWF), the EU network SCALA, the European Research Council (ERC)
and of the Institut f\"{u}r Quanteninformation Ges.m.b.H. 
\section{Correspondence}
\noindent Correspondence should be addressed to R.B.~(email:
\mbox{rainer.blatt@uibk.ac.at}).
\end{addendum}


\end{document}